# Robust Kagome Electronic Structure in Topological Quantum Magnets $X\text{Mn}_6\text{Sn}_6$ ($X$ = Dy, Tb, Gd, Y)

X. Gu[1*], C. Chen[2,3,4*], W. S. Wei[5], J. Y. Liu[6], X. Du[1], D. Pei[6], J. S. Zhou[1], R. Z. Xu[1], Z. X. Yin[1], W. X. Zhao[1], Y. D. Li[1], C. Jozwiak[4], A. Bostwick[4], E. Rotenberg[4], D. Backes[7], L. S. I. Veiga[7], S. Dhesi[7], T. Hesjedal[7], G. van der Laan[7], H. F. Du[5], W. J. Jiang[1,8], Y. P. Qi[2], G. Li[2,3], W. J. Shi[9,10], Z. K. Liu[2,3‡], Y. L. Chen[1,2,3,6‡], and L. X. Yang[1,8‡]

[1]State Key Laboratory of Low Dimensional Quantum Physics, Department of Physics, Tsinghua University, Beijing 100084, China.
[2]School of Physical Science and Technology, ShanghaiTech University and CAS-Shanghai Science Research Center, Shanghai 201210, China.
[3]ShanghaiTech Laboratory for Topological Physics, Shanghai 200031, China.
[4]Advanced Light Source, Lawrence Berkeley National Laboratory, Berkeley, California 94720, USA
[5]Anhui Province Key Laboratory of Condensed Matter Physics at Extreme Conditions, High Magnetic Field Laboratory of the Chinese Academy of Sciences, Hefei 230026, People's Republic of China
[6]Department of Physics, Clarendon Laboratory, University of Oxford, Parks Road, Oxford OX1 3PU, UK.
[7]Diamond Light Source, Harwell Science and Innovation Campus, Didcot OX11 0DE, United Kingdom
[8]Frontier Science Center for Quantum Information, Beijing 100084, China.
[9]Center for Transformative Science, ShanghaiTech University, Shanghai, 201210, China
[10]Shanghai High Repetition Rate XFEL and Extreme Light Facility (SHINE), ShanghaiTech University, Shanghai, 201210, China

\* These authors contributed equally to this work.

‡Email address: LXY: lxyang@tsinghua.edu.cn, YLC: yulin.chen@physics.ox.ac.uk, ZKL: liuzhk@shanghaitech.edu.cn

**Crystal geometry can greatly influence the emergent properties of quantum materials. As an example, the kagome lattice is an ideal platform to study the rich interplay between topology, magnetism, and electronic correlation. In this work, combining high-resolution angle-resolved photoemission spectroscopy and *ab-initio* calculation, we systematically investigate the electronic structure of $X\text{Mn}_6\text{Sn}_6$ ($X$ = Dy, Tb, Gd, Y) family compounds. We observe the Dirac fermion and the flat band arising from the magnetic kagome lattice of Mn atoms. Interestingly, the flat band locates in the same energy region in all compounds studied, regardless of their different magnetic ground states and 4$f$ electronic configurations. These observations suggest a robust Mn magnetic kagome lattice across the $X\text{Mn}_6\text{Sn}_6$ family, thus providing an ideal platform for the search and investigation on new emergent phenomena in magnetic topological materials.**

Kagome lattice, a two-dimensional network of corner-sharing triangles, exhibits many novel emergent properties such as spin liquids in frustrated magnets [1-3], competition between superconductivity and charge-density-waves [4,5], and interplay between magnetism, topology, and electron correlation [6-12]. Particularly, the destructive interference of Bloch electrons gives rise to flat bands that can be described by Chern numbers, mimicking the Landau levels without external magnetic field [13]. When the flat bands are partially filled, fractional quantum Hall state can be naturally realized [14,15]. In addition, as the kagome lattice is structurally a sibling of the honeycomb lattice, Dirac fermions exist at the corners of the Brillouin zone (Fig. 1a), similar to the physics in graphene [16]. If the time-reversal symmetry is broken, e. g., by ferromagnetic ordering, the band degeneracy of the Dirac cones will be lifted and a Chern gap will be induced as a result of the spin-orbit coupling (Fig. 1b), leading to the quantum anomalous Hall effect [17-19].

While the kagome model considers $s$ electrons and nearest-neighbor interaction, the situation is much more complex in real magnetic materials. On the one hand, many magnetic kagome materials involve $3d$ electrons with strong electron correlation; on the other hand, the kagome lattices in real materials usually consist of more than one type of atoms and may not be the ideal form due to their three-dimensional (3D) or quasi-2D crystal structure, making the realization of proposed topological electronic structures challenging. Although some intriguing phenomena owing to the Berry curvature of their topological electronic structures, such as the large anomalous Hall effect, have been observed in several kagome magnets [7,10,20-22], direct visualization of the massive Dirac fermions is still limited. Moreover, the characteristic flat bands near the Fermi level ($E_F$) are mainly observed in paramagnetic kagome materials so far, while in ferromagnetic kagome materials, the flat bands usually reside well below (> 200 meV) the Fermi level ($E_F$) [9,12,23-26], making the interplay between the magnetism and kagome-related physics less straightforward. Recently, it was proposed that there exists ideal kagome lattice with quantum-limit Chern topological magnetism in $X$Mn$_6$Sn$_6$ ($X$ = Dy, Tb, Gd, Y) family compounds, where the kagome layers are constructed solely by the Mn atoms. The existence of massive Dirac fermion at the $K$ point has been indicated by the observation of Landau quantization and Landau fan diagram in TbMn$_6$Sn$_6$ [7], and the flat band together with the Dirac cone have been directly observed in YMn$_6$Sn$_6$ and GdMn$_6$Sn$_6$ [26,27]. Nevertheless, the systematic evolution of the electronic structure of $X$Mn$_6$Sn$_6$ family and the influence of the magnetism with variable $X$ atoms still awaits further experimental exploration.

In this work, we systematically investigate the electronic structures of $X$Mn$_6$Sn$_6$ (X = Dy, Tb, Gd, Y) family compounds combining the use of high-resolution angle-resolved photoemission spectroscopy (ARPES) and *ab-initio* calculation (methods in the Supplemental Material [28]). Our experiment reveals multi-band nature of the electronic structure near $E_F$ derived from Mn $3d$ orbitals that shows clear $k_z$ dispersion, suggesting the important role of inter-kagome-layer coupling. We also observe characteristic electronic structures of Mn kagome lattice: the Dirac fermion near $E_F$ at the $K$ point and the flat bands, consistent with our *ab-initio* calculations. Remarkably, the observed kagome band structures in different compounds are almost identical, suggesting a minor influence of the magnetic ordering and $f$ electron configuration on the electronic structure of Mn kagome layer. These observations not only provide insights into the interplay between magnetism, crystal geometry, correlation, and topology, but also shed light on the future study of novel emergent phenomena in magnetic topological materials.

$X$Mn$_6$Sn$_6$ crystallize into a hexagonal structure with the space group of *P*6/mmm. Each unit cell contains a $X$-Sn layer and two Mn-Sn layers separated by a honeycomb layer of Sn atoms. Within each Mn-Sn layer, the Mn atoms form a two-dimensional kagome lattice (Fig. 1c), providing an ideal platform for the study of kagome-related emergent physics. At room temperature, the magnetic moments of the Mn atoms align ferromagnetically within each kagome layer, while the $X$ atoms engineer the overall magnetic ground state of the system [29-31] (Fig. 1d). In the compounds with $X$ = Dy, Tb, and Gd, adjacent Mn kagome bilayers couple ferromagnetically across the magnetic $X$-Sn layer, inducing ferrimagnetism with different magnetic anisotropies in the system [29,30], as shown in Fig. 1(d). In YMn$_6$Sn$_6$, by contrast, the Y atom exhibits no magnetization thus the adjacent ferromagnetic kagome bilayers couple antiferromagnetically across the non-magnetic Y-Sn layer, inducing complex helical antiferromagnetism below 300 K [32]. Details of the magnetism of $X$Mn$_6$Sn$_6$ compounds can be found in the Supplemental Material [28].

Figure 2 shows the electronic structure of a representative compound, DyMn$_6$Sn$_6$ measured at 12 K (no change of the magnetism of $X$Mn$_6$Sn$_6$ compounds is observed below 100 K. Supplementary Material, Fig. S1 [28]). We observe clear $k_z$ variation of the electronic states near $E_F$ (Supplemental Material, Fig. S3 [28]), as shown in the Fermi surface (FS) in the $k_z$-$k_\parallel$ plane in Fig. 2(b), indicating the strong inter-layer coupling (the inner potential used in the estimation of $k_z$ values is 10 eV). Fig. 2(c) shows the evolution of the constant energy contours with the binding energy at the $\Gamma KM$ plane (measured with 138 eV photon energy). The FS consists of a warped electron pocket around the $\Gamma$ point and triangle features near the $K$ point, which gradually evolves into a dot and corner-connected triangles respectively with increasing binding energy. Fig. 2(d) presents the 3D plot of the band structure showing sharp dispersions crossing $E_F$ along both the $\Gamma M$ and $\Gamma K$ directions with clear anisotropy. Fig. 2(e) and 2(f) show the band dispersions along high-symmetry directions measured with different photon polarizations. In linear-horizontal (LH) channel, we observe two bands (the inner $\alpha$ band and outer $\beta$ band) crossing $E_F$ around $\Gamma$. The $\alpha$ band crosses $E_F$ along both $\Gamma K$ and $\Gamma M$, forming the warped FS around $\Gamma$, while the $\beta$ band forms a Dirac-cone near $E_F$ at the $K$ point, which is the characteristic feature of the kagome lattice. In linear-vertical (LV) channel, except for the $\beta$ band observed in LH channel, two other bands crossing $E_F$ ($\gamma$ and $\delta$) are evidenced. Interestingly, the $\delta$ band has an electron-like dispersion along $KM$ while all the bands show hole-like dispersion along $M\Gamma$. Therefore, the $\delta$ band forms a saddle point near the $M$ point (Supplemental Material, Fig. S5 [28]), which is another characteristic feature anticipated in the kagome lattice. In Fig. 2(g), the calculated band structure, after being renormalized by a factor of 2.75 and shifted by 35 meV, is superposed with experimentally extracted band dispersions around the Dirac point at *K*. The observed band structure near the Dirac point can be well captured by the calculation with spin-orbit coupling included, which predicts a nearly gapless Dirac cone (energy gap < 5 meV) near $E_F$. Noticeably, our calculation suggests another Dirac cone with a large energy gap (~ 30 meV) at 100 meV above $E_F$, in good agreement with the simulation of the Landau fan data [7].

To further investigate the influence of the *X* atoms on the electronic structure, we compare the electronic structures of different $X$Mn$_6$Sn$_6$ compounds, as illustrated in Fig. 3(a-d). The FS structure in general does not change with the variation of *X* element, even for the non-magnetic Y atoms, suggesting the weak influence of the *X* element and the robustness of the Mn kagome lattice. Consistently, similar band dispersions along Γ*KM* are observed in all the four compounds studied [Fig. 3(e-h)], with only a small variation of the Fermi momenta of the α band, possibly induced by a slight change of the doping level with different *X* atoms.

Our *ab-initio* calculation indicates the existence of flat bands around 0.4 eV below $E_F$ that originate from Mn $d_{z^2}$ orbitals (Fig. 2(g) and Supplemental material, Fig. S5 [28]). To search for these flat bands, we conducted photon-energy dependent ARPES measurements over a large energy range (60 to 200 eV), with the results demonstrated in Fig. 4. Due to the orbital characters of the flat bands, they are better visualized using LV photons (Supplemental Material, Fig. S8 [28]). Fig. 4(a) shows the ARPES spectrum along the Γ*K* direction measured on DyMn$_6$Sn$_6$ using 112 eV photons, and as expected, the flat band is clearly observed at 0.42 eV below $E_F$. Fig. 4(b) and 4(c) illustrate the result of the same measurements on TbMn$_6$Sn$_6$ and GdMn$_6$Sn$_6$, and similar flat bands are ubiquitously observed. The flat band can also be clearly seen in the measurements along $k_z$ direction at different $k_y$ values [Fig. 4(d-f)], with minor differences among different compounds. In addition, a second flat band is observed around -100 meV for all compounds as indicated by the red arrows in Fig. 4(a-c), which was likewise observed in the extended Brillouin zone in $YMn_6Sn_6$ [26].

Despite the change of the $X$ atoms and the magnetism configuration of the system, the kagome band structure of $XMn_6Sn_6$ remains almost the same (Fig. 4, also see Supplementary Material, Fig. S4 and S7 [28]), suggesting a robust and ideal platform to investigate and engineer the kagome-related physics. Nevertheless, we notice that the flat bands are not observed over the whole Brillouin zone and show weak dispersion with a band width of about 50 meV (Fig. 4), which may result from the strong inter-kagome-layer coupling. Quantum-limit kagome physics is expected in the atomically thin MnSn layer if it can be isolated from $XMn_6Sn_6$ or synthesized by molecular beam epitaxy. Moreover, if the Fermi level can be tuned into the large gap of the Dirac cone at about 100 meV above $E_F$ [Fig. 2(f)], interesting quantum effect may be realized.

Besides, the $XMn_6Sn_6$ compounds provide an ideal platform to study the interplay between correlation and topology [27]. The correlation effect usually plays an important role in the $3d$ electronic systems, manifested here in different ways in $XMn_6Sn_6$ system. Firstly, a moderate band renormalization factor of about 2.75 has to be used to compare the calculated and observed band structure. Secondly, highly dispersive bands with band width larger than 1 eV are observed [shown in Fig. 3(e-h)], in drastic contrast to the calculation but reminiscent of the waterfall dispersions in cuprates and iridates [33-37], indicating the impact of electronic correlation on the electronic structure of $XMn_6Sn_6$. Thirdly, the bands below 0.5 eV are strongly blurred and/or disappear in our experiment, suggesting that the electronic states are incoherent at high binding energies, which is commonly observed in correlated materials [38-40]. Finally, the $4f$ electrons of the rare earth elements Dy, Tb, and Gd seem to not strongly interact with Mn $3d$ electrons, and their influence on the flat bands of the kagome layer is negligible. Nevertheless, the waterfall dispersion is suppressed and blurred in rare-earth compounds compared to that in $YMn_6Sn_6$ (Fig. 3), suggesting that the $4f$ electrons do contribute to the correlation effect in $XMn_6Sn_6$. The complex correlation effect requires further experimental and theoretical investigation to understand its interplay with the magnetism and topology in these kagome materials.

In conclusion, we have systematically studied the electronic structure of magnetic kagome materials $X$Mn$_6$Sn$_6$ and observed the Dirac cone, the saddle point, and flat bands that are characteristic features of the electronic structure of kagome lattice. The observed electronic structure, particularly the flat bands, is robust against the change of magnetic moment and 4$f$ electronic configuration related to the $X$ atoms. Our results not only provide important insights into the electronic structure of an ideal family of magnetic kagome materials, but also suggest an interesting platform to investigate novel topological kagome physics and its interplay with magnetism and correlation effect. We further propose that rich and intriguing phenomena are expected in the atomically thin films of the Mn kagome layer.

**Acknowledgment**

This work was supported by the National Natural Science Foundation of China (Grants No. 11774190, No. 11674229, No. 11634009), the National KeyR&D program of China (Grants No. 2017YFA0304600 and No. 2017YFA0305400), EPSRC Platform Grant (Grant No. EP/M020517/1). W. Shi acknowledges the support from Shanghai-XFEL Beamline Project (SBP) (31011505505885920161A2101001). The calculations were carried out at the Scientific Data Analysis Platform of Center for Transformative Science and the HPC Platform of ShanghaiTech University Library and Information Services. This research used resources of the Advanced Light Source, a U.S. DOE Office of Science User Facility under contract no. DE-AC02-05CH11231. Diamond light source is acknowledged for beamtime at beamline I06 under proposal number MM27482.

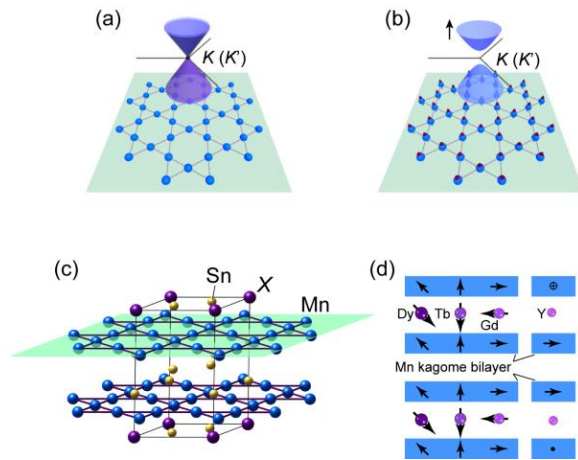

FIG. 1.(a) Non-magnetic kagome lattice with a gapless Dirac cone at the K point of the Brillouin zone (BZ). (b) Ferromagnetic kagome lattice with a gapped Dirac cone at K. (c) Crystal structure of $X\mathrm{Mn}_6\mathrm{Sn}_6$ ($X$ = Dy, Tb, Gd, Y). The green plane highlights the kagome layer of Mn atoms. (d) Schematic of the magnetism of $X\mathrm{Mn}_6\mathrm{Sn}_6$ at low temperatures. The kagome layers of Mn atoms show different ferromagnetism for rare-earth materials, while the Mn layers in $\mathrm{YMn}_6\mathrm{Sn}_6$ show a helical antiferromagnetism. The magnetization of the $X$ atoms changes in different compounds.

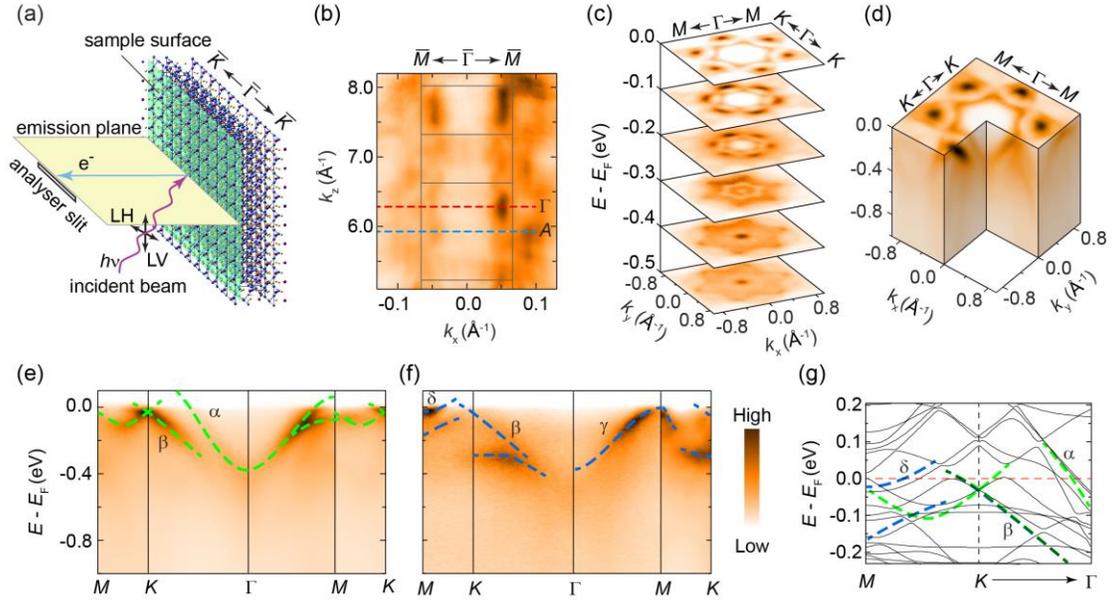

FIG. 2. (a) ARPES geometry used in our experiments. (b) Fermi surface (FS) of DyMn$_6$Sn$_6$ in the $k_z$-$k_x$ plane measured with photon-energy dependnent ARPES experiment. (c) Evolution of the constant energy contours with binding energy measured with 138 eV photon energy. (d) 3D plot of the electronic structure on the $k_x$-$k_y$ plane. (e), (f) Band structure along high-symmetry directions measured with linear-horizontally (LH, d) and linear-vertically (LV, e) polarized photons at 138 eV. The green and blue curves are guide to the eyes for the measured dispersion. Note that the Fermi momenta of the α and γ bands are different. (g) Comparison of the measured and calculated band structure along high symmetry directions. The black curves are the calculated band structure after being renormalized by a factor of about 2.75 and shifted by 35 meV. Data were collected at 12 K.

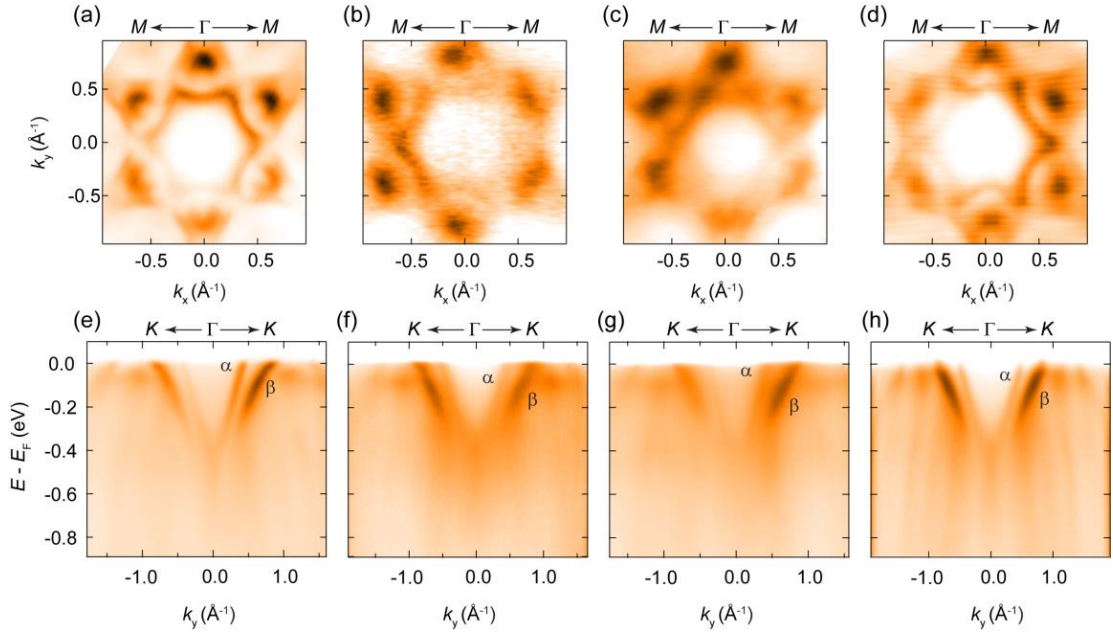

FIG. 3. (a-d) FS of (a) DyMn$_6$Sn$_6$, (b) TbMn$_6$Sn$_6$, (c) GdMn$_6$Sn$_6$, and (d) YMn$_6$Sn$_6$. (e-h) Band dispersion along *MK*Γ in different compounds. Data were collected using linear-horizontally polarized photons at 138 eV at 12 K.

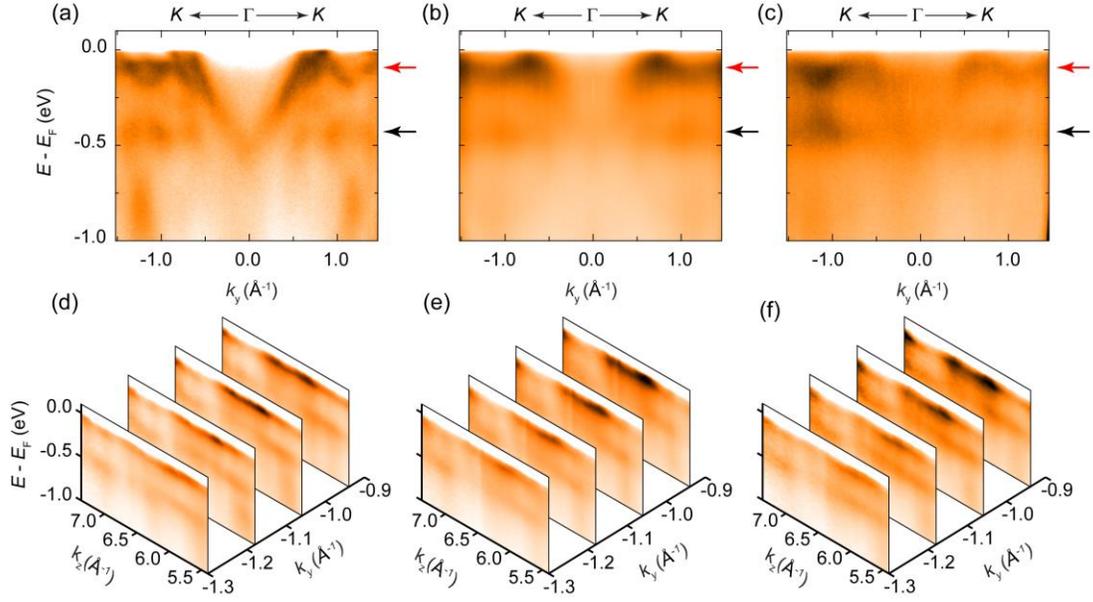

FIG. 4. (a-c) Band dispersion of (a) DyMn$_6$Sn$_6$, (b) TbMn$_6$Sn$_6$, and (c) GdMn$_6$Sn$_6$ along MKΓ. (d-f) Band dispersions of $X$Mn$_6$Sn$_6$ along $k_z$ at different $k_y$. (a-c) are the summation of the data measured with linear-vertically and linear-horizontally polarized photons at 112 eV. Data in (d-f) were collected using linear-horizontally polarized photons. All data were collected at 12 K.